\documentclass[letterpaper,12pt,oneside,onecolumn,final]{article}
\input{epsf}

\begin{document}

\author{E. Cuautle$^{ac}$, \and G. Herrera$^a$, \and J. Magnin$^b$
\and $\underline{\mbox{A. S\'anchez-Hern\'andez}}^a$
\thanks{ $\mbox{e-mails:
cuautle@lafex.cbpf.br, gherrera@fis.cinvestav.mx,}$
\hspace*{1.9cm}jmagnin@invefis1.uniandes.edu.co, asanchez@fis.cinvestav.mx.}
\\{\small $^a$Centro de Investigaci\'on y de Estudios Avanzados}\\
  {\small Apdo. Postal 14 740, M\'exico D.F 07000, M\'exico}
\\{\small $^b$Departamento de  F\'{\i}sica, Universidad de los Andes}\\
  {\small A.A. 4976, Bogot\'a D.C. Colombia}
\\{\small $^c$Centro Brasileiro de Pesquisas F\'{\i}sicas}\\
  {\small Rua Dr. Xavier Sigaud 150, Rio de Janeiro 22090-150 RJ, Brasil}
}

\title{Charge asymmetries in charm photoproduction
\thanks{ This work has been supported by CONACyT, ColCiencias, and CLAF--CNPq}}


\maketitle

\begin{abstract}
Charm quarks are expected to be produced mainly by the photon-gluon 
fusion mechanism in $\gamma-N$ interactions. However, a small part 
of the total charm cross section originates in similar processes 
to those appearing in charm hadroproduction through the resolved 
(hadronic) component of the photon. Although the contribution of the 
resolved part of the photon is small at  fixed target energies, it can 
help to understand the small but sizeable charge asymmetries 
measured in charm hadron photoproduction experiments.
\end{abstract}

\section{Introduction}

Charm hadrons in photon nucleus interactions at typical fixed target 
experiment energies, are expected to be produced predominantly by 
photon-gluon fusion followed by fragmentation.

As $c$ and $\bar{c}$ quarks in the $\gamma g \rightarrow c \bar{c}$ process 
are produced at the same rate, apart from a tiny $\bar{c}$ excess appearing 
from Next to Leading Order (NLO) contributions~\cite{nason-ellis}, the 
final charm hadron and anti-hadron cross sections should be approximately 
the same. Furthermore, associated production, which together with 
Leading Particle Effects (LPE) has been observed 
to play an important role in the hadroproduction asymmetries for charm and 
anti-charm particles~\cite{leading,eleazar}. This should not induce charge 
asymmetries in 
photoproduction since the effects are the same for particles than for 
anti-particles within this scheme.

However, the SLAC Hybrid Photon Facility Collaboration~\cite{slac} has 
reported a noticeable charge asymmetry in meson photoproduction. This result 
has been confirmed more recently by the E691~\cite{e691} and the 
E687~\cite{e687} charm photoproduction experiments. Actually, the E687 
Collaboration presented results on charge asymmetries in $D^-/D^+$ and 
$D^{*-}/D^{*+}$ photoproduction which are consistent, at a three sigma 
level, with a positive asymmetry. Results on $D_s^-/D_s^+$, 
$\bar{D}^0/D^0$ and even $\Lambda_c^+/\Lambda_c^-$ 
photoproduction asymmetries, from both the E687 and the E691 experiments, 
are less clear since the error bars are still large to be conclusive, but 
all these measurements seems to indicate a small charge asymmetry (See 
Table~\ref{tab1} for a summary of the E687 and E691 results).

\begin{table}[t]
\begin{center}
\begin{tabular}{|c|c|c|c|} \hline \hline
Particle & Decay mode & $R$ (E691)~\cite{e691} & $R$ (E687)~\cite{e687}    \\ \hline
$D^0$    & $K^-\pi^+$                    & $1.08\pm0.03$    & $1.04\pm0.03$ \\
         & $K^-\pi^+\pi^+\pi^-$          &                  & $1.04\pm0.03$ \\
$D^+$    & $K^-\pi^+\pi^+$               & $1.04\pm0.3$     & $1.08\pm0.02$  \\
$D^{*+}$ & $D^0\pi^+$                    & $1.15\pm0.07$    & $1.13\pm0.03$  \\
         & ($D^0\rightarrow K^-\pi^+$)   &                  &                \\
$D^{*+}$ & $D^0\pi^+$                    & $1.23\pm0.07$    & $1.08\pm0.04$  \\
         & ($D^0\rightarrow K^-\pi^+\pi^+\pi^-$) &          &                \\
$D_s^+$  & $\phi\pi^+ + \bar{K}^{*0}K^+$ &                  & $0.92\pm0.14$  \\
         & $K^-K^+\pi^+$                 &                  & $0.95\pm0.09$  \\
$\Lambda_c^+$ & $pK^-\pi^+$              & $0.79\pm0.17$    & $0.93\pm0.14$  \\
\hline \hline
\end{tabular}
\end{center}
\caption{$R=$ antiparticle/particle ratio. 
The $D^+/D^-$ was the statistically most significant sample of the 
E691 experiment (they saw a raw signal of $4864\pm103$ events, 
see Ref.~\cite{e691}). Results of the E687 Collaboration were 
taken from Ref.~\cite{e687}.}
\label{tab1}
\end{table}

The origin of the charge asymmetry in photoproduction remains unexplained. 
However, in Ref.~\cite{slac}, a simple model which qualitatively might 
account for the observed results has been presented. In this simple model, 
a charge asymmetry arises when a light anti-quark ($\bar{u}$ 
or $\bar{d}$) in the photon structure annihilates with a quark of the 
same flavor from the nucleus in the process 
$q\bar{q} \rightarrow q'\bar{q}'$, favoring thus the production 
of a particle containing the accompanying quark from the photon liberated 
in the interaction. This mechanism is twofold. On one side an asymmetry 
appears in the particle / anti-particle production due to the 
different content of quark and anti-quarks in the target nucleons and, on the 
other side, an asymmetry between final particles containing $u$ or $d$ 
valence quarks should arises because of the different content of 
$u$ and $d$ valence when targets are made of a different number 
of protons and neutrons.

	In Ref.~\cite{e687} a model based on the Lund--string fragmentation 
scheme is presented. In the model the color field between the target diquark
and the charm quark produced in the photon--gluon interaction build a string.
Similarly the bachelor quark build a string with the anticharm quark.
The model is discussed there and further details can be 
found in Ref.~\cite{lund}. A very good agreement  with experimental data is 
obtained with the two versions describe there.

	Here we try to understand the observed asymmetry in terms of the 
effects introduced by the resolved component of the photons.

\section{Charm photoproduction cross sections}

The invariant cross section for the photoproduction of a heavy quark is 
as follow~\cite{nason-ellis}
\begin{equation}
\frac{Ed^3\sigma}{dp^3} = 
\sum_i{\int{dx \frac{Ed^3\hat{\sigma}_{\gamma i}}{dp^3} f^H_i(x)}} + 
\sum_{i,j}{\int{dx_1dx_2 \frac{Ed^3\hat{\sigma}_{i,j}}{dp^3} f^{\gamma}_i(x) 
f^H_j(x)}} \; .
\label{eq1}
\end{equation}

In eq.(\ref{eq1}), a $\mu$ dependence is implicit in the 
elementary cross sections $\hat{\sigma}$ and the number densities 
of light partons (gluon, light quarks and anti-quarks) in the hadron 
($f^H_i(x)$) and the photon ($f^{\gamma}_i(x)$). The short distance 
cross sections $\hat{\sigma}$ are calculable as a perturbative 
series in $\alpha(\mu^2)$. 

The first term in the right hand side of eq.(\ref{eq1}) known as 
point like contribution, while the second term is the hadronic 
component of the photon. The separation of the two terms is controlled 
by the scale $\mu$ (see Ref.~\cite{nason-ellis} for a detailed discussion).

Typical contributions to the first term of eq.(\ref{eq1}) are
\begin{eqnarray}
\gamma + g  &\rightarrow&  c + \bar{c} \nonumber \\
\gamma + g  &\rightarrow&  c + \bar{c} + g \nonumber \\
\gamma + q  &\rightarrow&  c + \bar{c} + q \nonumber \\
\gamma + \bar{q}  &\rightarrow&  c + \bar{c} + \bar{q}
\label{eq2}
\end{eqnarray}
where the first process receives contributions from Leading and Next to 
leading order while the second and following appear only at NLO. 
To this order in the perturbative series of the point like photon 
coupling, a tiny difference arises in the $c$ and $\bar{c}$ cross sections. 
However, this effect is very small to account for the observed 
asymmetries in charm meson photoproduction.

The partonic subprocesses contributing to the resolved photon 
component up to NLO are
\begin{eqnarray}
g+g \rightarrow c + X &\hspace{2cm}&  g+g \rightarrow \bar{c} + X \nonumber \\
g+q \rightarrow c + X &\hspace{2cm}&  g+q \rightarrow \bar{c} + X \nonumber \\
g+\bar{q} \rightarrow c + X &\hspace{2cm}& g+\bar{q} \rightarrow \bar{c} + X \nonumber \\
q+\bar{q} \rightarrow c + X &\hspace{2cm}& q+\bar{q} \rightarrow \bar{c} + X
\label{eq3}
\end{eqnarray}
where $X$ is either a $c$ quark (anti-quark) or a $c$ quark (anti-quark) 
plus a gluon. All these terms receive contributions from both Leading 
and Next to Leading Order (see Ref.~\cite{nason-ellis2}). Again, 
at NLO a tiny excess of $\bar{c}$ over $c$ quarks appears.

\begin{figure}[htb]
\epsfysize=7.0cm
\centerline{\epsffile{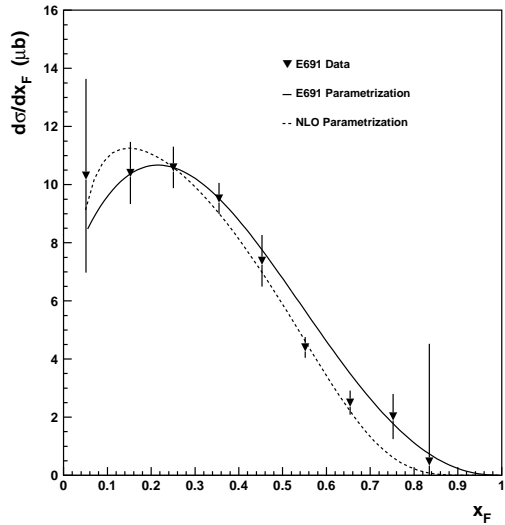}}
\caption{$D^{\circ}$ and $D^{\pm}$ differential cross sections as a 
function of $x_F$ compared to experimental data from the E691 
Collaboration~\cite{e691}. For the NLO parametrization we have used
GRV-LO and GRVG-LO~\cite{grvs} parton distribution functions for hadron and 
photon respectively.}
\label{fig1}
\end{figure}

The convolution of the differential cross section $d\sigma/dx_F$ for charm 
quark production with the Peterson fragmentation function 
\begin{equation}
D_{c/H_c}(z) = \frac{N}{z\left[1 - \frac{1}{z} - 
\frac{\epsilon}{1-z} \right]^2} \; ,
\label{eq4}
\end{equation}
gives a good description of the main features of $D^{\circ}$ and 
$\bar{D}^{\circ}$ photoproduction with $\epsilon = 0.11$, and 
$D^{\pm}$ photoproduction with $\epsilon = 0.09$ (See Fig.~\ref{fig1}). 
However, as fragmentation is the same for $c$ than for $\bar{c}$, no 
further asymmetries than those already appearing from the NLO 
calculation of heavy quark production arise in the final hadronic state.

\section{The hadronic contribution of the photon}

\begin{figure}[htb]
\epsfysize=7.0cm
\epsfxsize=5.5in
\centerline{\epsffile{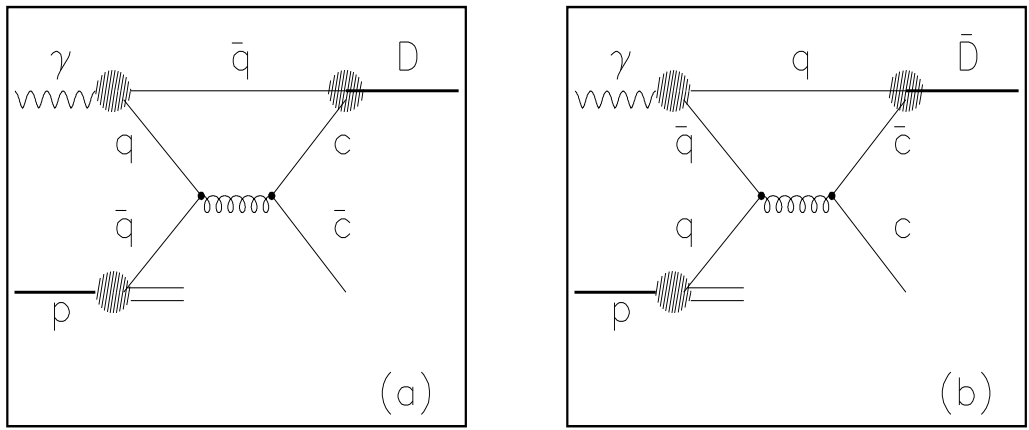}}
\caption{Production of charm mesons  and anti-mesons 
in the model. The cross section of the process in (a) must be smaller than the 
one in (b) just because antiquarks density in photons is smaller than quarks 
densities.}
\label{fig4}
\end{figure}

The hadronic contribution to the total cross section of eq.(\ref{eq1}) 
can produce additional contributions to the charm hadron production 
asymmetry {\em via} the mechanism outlined in Ref.~\cite{slac}. 
When a resolved photon interacts with a nucleon in the target, the process 
$\bar{q}_{\gamma} q_{N} \rightarrow c + \bar{c}$ is favored over the 
process $q_{\gamma} \bar{q}_{N} \rightarrow c + \bar{c}$ due to the 
partonic structure of nucleons. Then an excess in the production of 
mesons containing a $c$ over mesons containing a $\bar{c}$ quarks 
arises at the hadronization level. This is due to the fact that the 
produced $\bar{c}$ quark can recombine easily with the $q_{\gamma}$ 
liberated in the collision to produce an anti-meson in the final state. 
The recombination of the $c$ quark with the $q_{\gamma}$ should instead 
produce a baryon. This mechanism tends to produce more charm anti-mesons than 
mesons and, conversely, more charm baryons than anti-baryons, 
as the experimental data seems to indicate. $D_s^{\pm}$ meson photoproduction 
should not present any asymmetry at all since $s=\bar{s}$ in both the 
nucleon and the photon. Fig.~\ref{fig4} shows a pictorical 
representation of the model.

Furthermore, as the proton has two $u$ and one $d$ valence quarks, 
the production of $\bar{D}^{\circ}\; (u\bar{c})$ should be favored 
over the production of $D^-\; (d\bar{c})$ in $\gamma-proton$ interactions. 
The opposite must happen in $\gamma-neutron$ interactions. Notice also that, as 
the results obtained by the E691 and E687 Collaborations are on 
$\gamma$-Beryllium interactions, and since the Beryllium nucleus has 
more neutrons than protons, and excess of $D^-$ over $\bar{D}^{\circ}$ 
should appear. Although none of the above experiments have measured 
the $D^-/\bar{D}^{\circ}$ asymmetry, predictions of the model seems to agree 
with the experimental data as long as the anti-particle to particle 
ratio is bigger for $D^-/D^+$ than for $D^-/\bar{D}^{\circ}$. 
Notice that the ratio $D^+/D^{\circ}$ should be approximately one, 
independently of the target particle since production in the forward 
($x_F>0$ region in these cases should 
proceeds mainly through independent fragmentation of the $c$ quark.

The hadronization of the perturbatively produced $c\,(\bar{c})$ quark 
through the recombination with the debris of the photon can be calculated 
along the lines developed in Ref.~\cite{kartvelishvili}. However, the 
recombination processes are not easy to be quantitatively estimated. 
On one hand, the momentum correlation between the perturbatively produced 
$c\,(\bar{c})$ quark and the fragments of the photon must be accounted for 
and, on the other hand, there exist an inherent difficulty associated with 
the definition of the multiquark distribution and the recombination 
function within this scheme.

Nevertheless, as the recombination process should not enhance the 
already present asymmetry, it is still possible to make some 
quantitative estimates which we shall present in the next section.

\section{Estimation of charge asymmetry from recombination}

In order to make a quantitative estimate of the charge 
asymmetry induced by diagrams in Fig.(\ref{fig4}), we define 
\begin{equation}
A = \frac{ \sigma_{q_{\gamma}\bar{q}_p} - \sigma_{\bar{q}_{\gamma}q_p}}
{2\sigma_{\gamma \; g} + \sigma_{H\; p}}
\label{eq5}
\end{equation}
where 
\begin{eqnarray}
\sigma_{\bar{q}_{\gamma}q_p} &=& \sum_{i,j}{\int{dx_1 dx_2 \;
\bar{q}^{\gamma}_i(x_1) \; q^{p}_j(x_2) \; E \; \frac{d^3\hat{\sigma}_{i,j}} 
{dp^3}}} \label{eq6a} \\
\sigma_{q_{\gamma}\bar{q}_p} &=& \sum_{i,j}{\int{dx_1 dx_2 \;
q^{\gamma}_i(x_1) \; \bar{q}^{p}_j(x_2) \; E \; \frac{d^3\hat{\sigma}_{i,j}} 
{dp^3}}}
\label{eq6b}
\end{eqnarray}
are the cross sections for the production of a $c-\bar{c}$ pair 
from light quark anti-quark annihilation. 
In eq.(\ref{eq6a}) the anti-quark 
comes from the photon while in (\ref{eq6b}) 
it originates in the proton structure.

$\sigma_{\gamma \; g}$ is the point like contribution to the total 
cross section of Eq.(\ref{eq1}) and $\sigma_{H\; p}$ is the sum 
of the cross sections of Eqs.(\ref{eq6a}), (\ref{eq6b}), and the 
gluon--gluon fusion processes appearing in the resolved photon contribution.

\begin{figure}[htb]
\begin{center}
  \begin{tabular}{cc}
    {\epsfxsize=2.5in 
     \epsffile{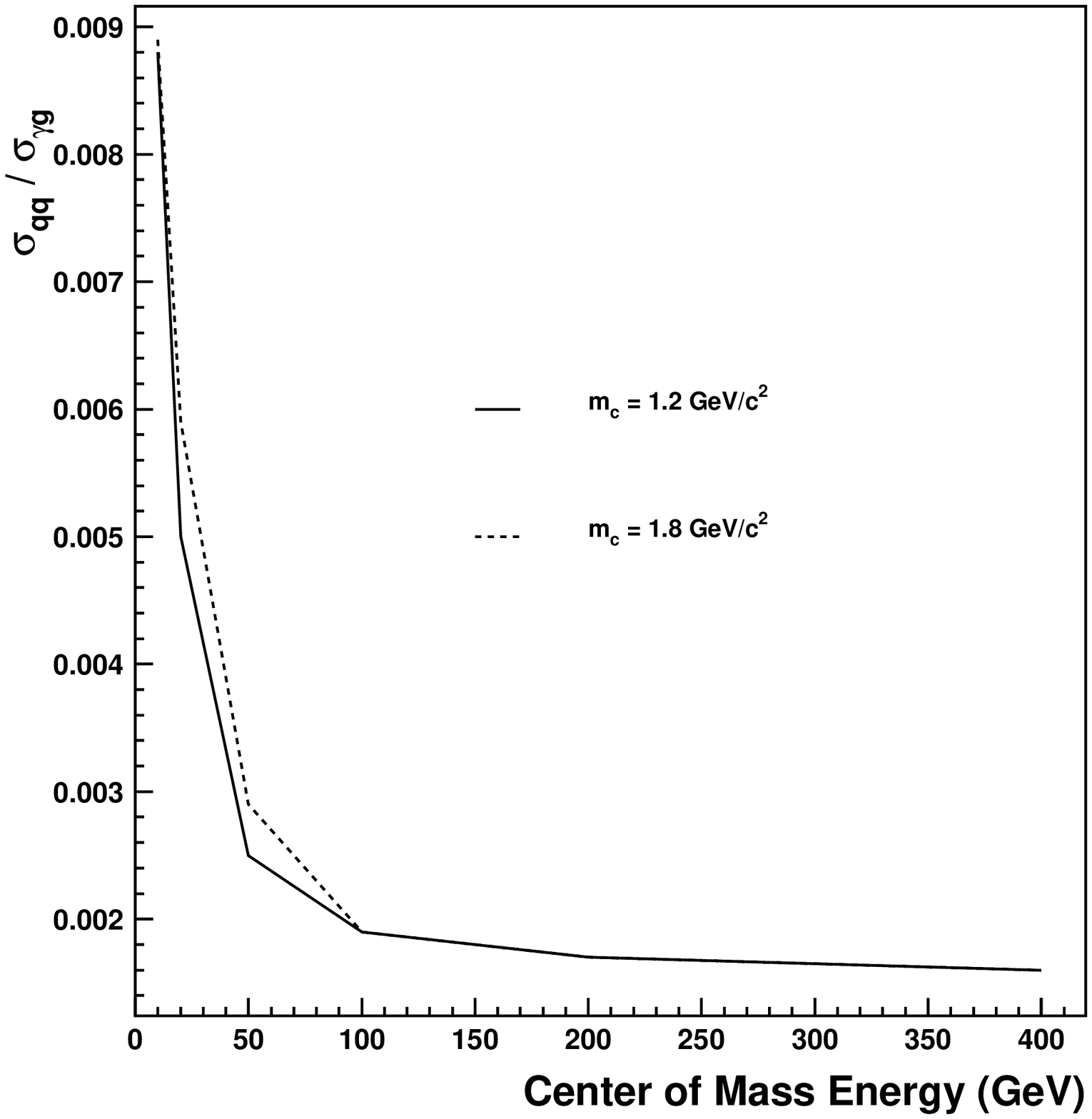}} & 
    {\epsfxsize=2.5in
     \epsffile{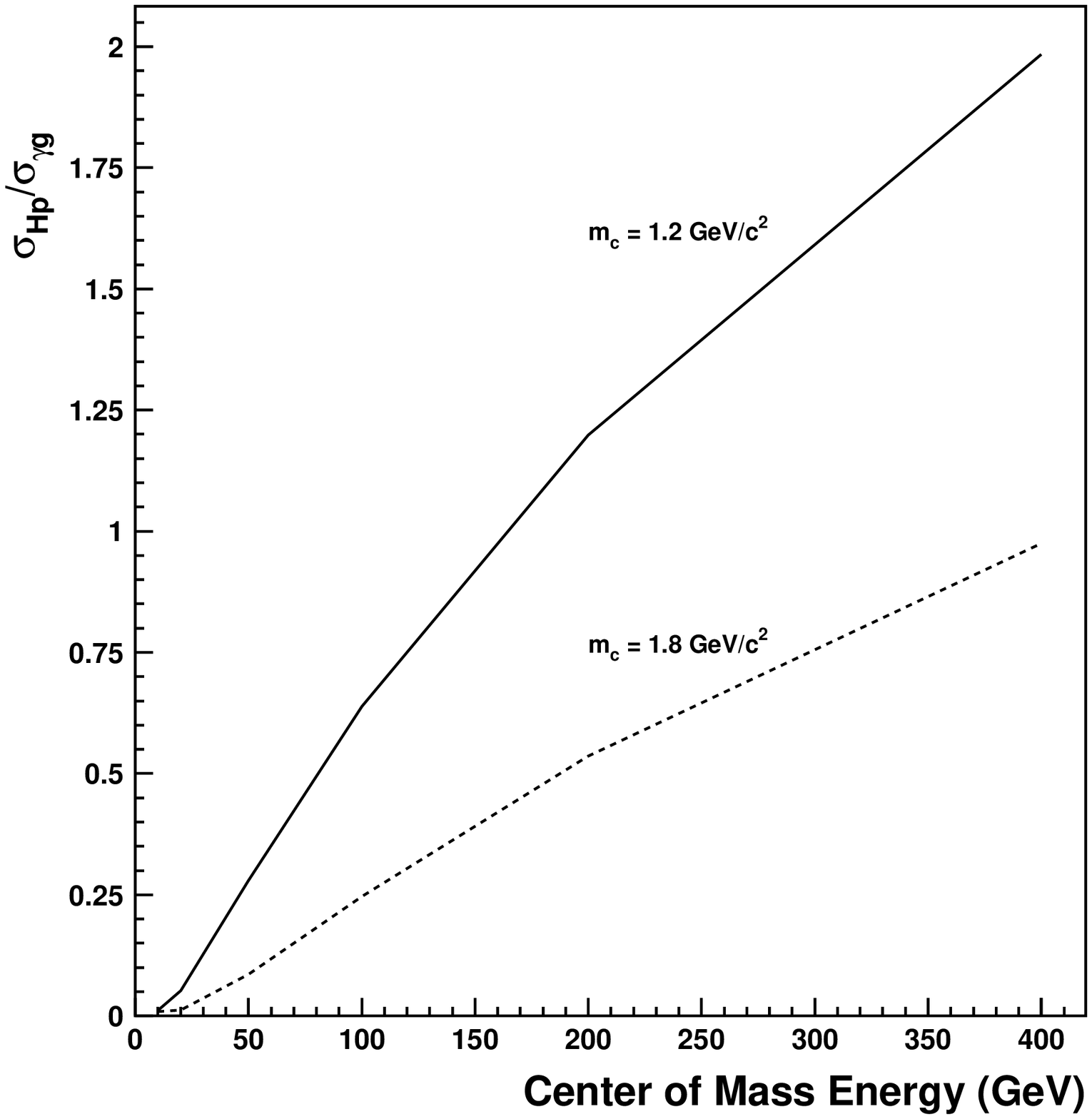}}
  \end{tabular}
  \caption{Contribution of each process to the total cross section 
for charm photoproduction as a function of the incident photon energy. 
$\sigma_{qq}$ represents the $\bar{q}_{\gamma} q_p$ and 
$q_{\gamma} \bar{q}_p$ processes.}
\label{fig5}
\end{center}
\end{figure}

As the separation of the resolved and point like contributions to 
eq.(\ref{eq1}) is controlled by the factorization scale entering in the 
photon and proton parton distributions, we have calculated the asymmetry 
defined by eq.(\ref{eq5}) for $\mu_F = 1;\; 2$ GeV. The renormalization 
scale has been also varied according to $\mu_R = a\; m_c$ with $m_c = 1.2;\; 
1.5;\; 1.8$ GeV and $a = \frac{1}{2};\; 1;\; 2$. The individual 
contributions of each process entering in eq.(\ref{eq5}) are plotted 
as a function of the incident photon energy for $\mu_F = \mu_R = m_c$ 
in Fig.~\ref{fig5},  the GRV-LO and GRVG-LO parton distribution functions 
have been used for calculations.

\begin{figure}[htb]
\epsfysize=7.0cm
\centerline{\epsffile{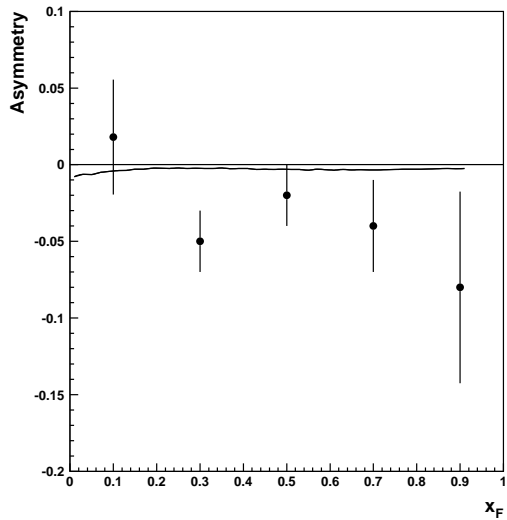}}
\caption{Total charge asymmetry. Dots represent experimental data from 
E687~\cite{e687}. The curve is
 the $c/\bar{c}$ asymmetry as a function of $x_F$ from 
NLO contributions, eq.(\ref{eq5}), 
in $\gamma-p$ interactions for a $200$ GeV energy beam.}
\label{fig2}
\end{figure}

The $c/\bar{c}$ asymmetry  appearing from NLO contributions to 
eq.(\ref{eq1}) is displayed in 
Fig.~\ref{fig2} for a $\gamma$ energy of $200$ GeV, the GRV-LO and GRVG-LO 
parton distribution functions 
were used. At this energy, 
the ratio of the $c$ to the $\bar{c}$ 
cross sections is 1.006, which is smaller than the $D/\bar{D}$ 
cross sections ratio measured in experiments (See Table~\ref{tab1}), but
within errors compatible with the more recent measurements of the E687 
experiment.

As can be seen in Fig.~\ref{fig5}, the 
$\bar{q}_{\gamma} q_p \rightarrow c+\bar{c}+X$ and 
$q_{\gamma} \bar{q}_p \rightarrow c+\bar{c}+X$ become less 
important as the photon energy rises up, indicating that any charge 
asymmetry arising from the resolved photon component must decrease 
with the photon energy, $E_{\gamma}$.

\section{Conclusions}

We have tried to explain the observed charge asymmetry in photoproduction
experiments using the resolved component of the photon. We have found that even
when this part is small for typical fixed target energies the charge asymmetry
raised from recombination is within errors consistent with the more recent
measurements of E687. The model of ref.~\cite{e687,lund} gives a larger
asymmetry than our approach but given the accuracy of experimental data it is
still hard to be conclusive on the responsible production mechanism.  We expect
that E831/FOCUS~\cite{e831} with its one million of charm reconstructed
candidates shed light on the issue.

\end{document}